# OCR Error Correction Using Character Correction and Feature-Based Word Classification

Ido Kissos
School of Computer Science, Tel Aviv University
Ramat Aviv, Israel

Nachum Dershowitz
School of Computer Science, Tel Aviv University
Ramat Aviv, Israel
Institut d'études avancées de Paris
Paris, France

*Abstract*—This paper explores the use of a learned classifier for post-OCR text correction. Experiments with the Arabic language show that this approach, which integrates a weighted confusion matrix and a shallow language model, improves the vast majority of segmentation and recognition errors, the most frequent types of error on our dataset.

## I. INTRODUCTION

A massive digitization of textual resources, such as books, newspaper articles and cultural archives has been underway for some decades, making these resources publicly available for research and cultural purposes. Institutions are converting document images into machine readable text via Optical Character Recognition (OCR), enabling a realistic way of exploring vast document corpora with automated tools, such as indexing for textual search and machine translation.

For reasons of low quality printing and scanning or physical deterioration, many of these corpora are of poor quality, making the OCR task notoriously difficult. Consequently, it is impossible to directly employ the obtained results for subsequent tasks without costly manual editing. Although contemporary OCR engines claim higher than 97% word accuracy for Arabic, for instance, the same datasets with low-resolution images or infrequent character classes can drop to lower than 70%.

We propose an OCR post-correction technique based on a composite machine-learning classification. The method applies a lexical spellchecker and potentially corrects single-error misspellings and a certain class of double-error misspellings, which are the major source of inaccurate recognitions in most OCR use-cases. The novelty of this method is its ability to take into consideration several valuable word features, each giving additional information for a possible spelling correction. It is built out of two consecutive stages:
1) word expansion based on a confusion matrix, and
2) word selection by a regression model based on word features.

The confusion matrix and regression model are built from a transcribed set of images, while the word features rely on a language model built from a large publicly available textual dataset.

The first stage generates correction candidates, ensuring high recall for a given word, while the second assures word-level precision by selecting the most probable word for a given position. Relying on features extracted from pre-existing knowledge, such as unigram and bigram document frequencies extracted from electronic dictionaries, as well as OCR metrics, such as recognition confidence and confusion matrix, we accomplished a significant improvement of text accuracy.

This research is part of the digitization project of the "Arabic Press Archive" of the Moshe Dayan Center at Tel Aviv University, hence the method evaluation and adaptation in in the Arabic language. There are a number of open-source and commercial OCR systems trained for this language; we used NovoDynamics NovoVerus commercial version 4, evaluated as one of the leading OCR engines for Arabic scripts.

We will use the Word Error Rate (WER) measure, which is appropriate for subsequent applications of the OCR output, such as information retrieval. Our correction method performs effectively, reducing faulty words by a rate of 35% on our dataset, which is a 10% absolute accuracy improvement. The overall results showed negligible false-positive errors, namely the method rarely rejects correct OCR words in favor of an erroneous correction, which is a major concern in spellcheckers. An analysis of classifier performance shows that bigram features have the highest impact on its accuracy, suggesting that the method is mainly context reliant.

This paper is organized as follows: Section 2 provides background information on Arabic OCR and OCR error correction. Section 3 presents the error correction methodology. Section 4 reports and discusses preliminary experimental results; and Section 5 concludes the paper and suggests some possible future directions.

## II. BACKGROUND

There is a significant amount of prior work on OCR error correction.

### A. The OCR Process

The goal of OCR is to extract the text, character by character, from a document image. The usual process proceeds in stages:

1) automatically segment a document image into images of individual characters in the proper reading order using image analysis heuristics;
2) apply an automatic classifier to determine the character codes that most likely correspond to each character image; and
3) exploit the surrounding context to select the most likely character in each position.

Preceding the OCR stage there is generally an imaging step, during which the image is prepared for character recognition, including, for example, document deskew, graphics and noise removal, etc. Following the OCR stage there can be a textual correction algorithm based on prior lexical knowledge of the document's language and subject domain.

*B. OCR Error Types*

OCR accuracy is negatively influenced by poor image quality (e.g., scanning resolution, noise) and any mismatch between the instances on which the character image classifier was trained and the rendering of the characters in the printed document (e.g., font, size, spacing). Depending on the language and the image quality of the analyzed collection, there will be a different error distribution generated by an OCR process. These errors can be categorized according to the following types, listed in the order they occur during the OCR process:

- Word detection – failing to detect text in the image, commonly caused by poor image quality or text mixed with graphics.
- Word segmentation – failing to bound an individual word correctly, due to wrong interword space detection, generally due to different text alignments and spacing.
- Character segmentation – failing to bound single characters in a segmented word. This is frequent for cursive or connected alphabets, such as printed Arabic or handwritten Latin-alphabet languages. It may also occur due to an analog process (e.g., printing and scanning speckles) that might disconnect connected components.
- Character recognition – failing to identify the correct character for a bounded character image.

*C. OCR Error Correction*

There has been much research aimed at the automated correction of recognition errors for degraded collections. An early, useful survey is [1]; relevant methods for Arabic OCR are summarized in [2] and in collection [3].

In this work, we use language models on the character and word levels, plus lexicons. We do not apply morphological or syntactical analyses, nor passage-level or topic-based methods.

Three language resources play a rôle:

- Dictionary lookup compares OCR-output with the words in a lexicon. When there is a mismatch, one looks for alternatives within a small edit (Levenshtein) distance, under the assumption that OCR errors are often due to character insertions, deletions, and/or substitutions. For this purpose, one commonly uses a noisy-channel model, a probabilistic confusion matrix for character substitutions, and term frequency lists [4], as we do here. One must, however, take into consideration unseen ("out of vocabulary") words, especially for morphologically-rich languages, like Greek, and even more so for *abjads*, like Arabic, in which vowels are not represented. The correct reading might not appear in the lexicon (even if it is not a named entity), while many mistaken readings will appear, because a large fraction of letter combinations form valid words. Morphological techniques could help here, of course. Dictionary lookup and shallow morphology are used in [2].
- We use the term $k$-mer[1] for the possible contiguous $k$-character substrings of words. By collecting statistics on the relative frequency of different $k$-mers for a particular language, one can often recognize unlikely readings. This technique was employed by BBN's OCR system for Arabic [5], as well as in [2].
- A language model, based on $n$-gram frequencies derived from a large corpus, is frequently used to estimate the likelihood of a reading in context [6].

III. ERROR CORRECTION METHODOLOGY

The OCR error model is vital in suggesting and evaluating candidates. At the heart of the error model is a candidate generation for correction, based on a confusion matrix giving conditional probabilities of character edits. The possible error corrections include the error types listed in Section II-B, except word detection problems as the algorithm has no input image to correct. The latter has to be addressed with image preprocessing or detection robustness.

We will focus the discussion at a word level at a certain position in the text, which is obtained by a standard tokenization of the OCR output text.

The error correction methodology comprises three stages:

1) Correction candidate generation – The original word is expanded by a confusion matrix and a dictionary lookup, forming all together a correction-candidates vector.
2) Feature extraction – Features are extracted for each word in the vector.
3) Word classification – A two-stage classification process, where the first stage ranks the correction candidates according to their correctness probability at this position, while the second selects the most probable between the original word and the highest-ranked candidate.

*A. Training Data*

The correction methodology is language, domain, scan quality and OCR engine agnostic; nevertheless the model itself is built upon a data corpus that resembles the test data.

Our experiment focused on OCR of printed Arabic documents of a commercial company archive. The archive documents are fairly variable, containing text in different fonts and

[1]Also known as "character $n$-gram". The term "$k$-mer", borrowed from bioinformatics, allows us to use "$n$-gram", unambiguously, for sequences of *words*.

page layouts, such as tables and graphics. We made use of the following training resources:

1) Two hundred fifty OCR document images and their ground truth transcription – each document is a A4 page scanned at 200 dpi. The relatively low-quality of scan was to emphasize the OCR correction ability to deal with low quality images and enrich the word errors the noisy channel training. The set was manually transcribed at a document level and OCR processed by Novodynamics Verus version 4.2. Fifty documents were left aside during the training process for later testing and evaluation. The training set contains about 60,000 words with a WER of 30%.
2) Company's digital archive – The corpus contains about 3 million documents from various Arabic documents. A modeling process produced unigram and bigram frequency lists, which were later used as features for correction candidate ranking and classification. Despite the corpus size, its frequency lists are not as accurate as they could be, as the content is not representative enough of the thematic domain of the same company's archive set. That is because the company's archive contains documents from distant expertise in contrast to the image documents.

### B. OCR Text Tokenizer

In order to structure the OCR text to enable correction at a word level, the text is tokenized by standard space delimiter tokenizer. This phase also parses the word recognition confidence produced by the OCR engine, forming a first level feature extraction.

### C. Correction-Candidate Generator

This module is designed to generate correction candidates for a tokenized word in accordance with an observed OCR error model. The error model we implemented supports the correction of erroneous character segmentation and recognition, as well as word segmentation. The former is handled by supporting primitive 1-Levenshtein distance[2] [7] plus 2:2 alignments, the latter by whitespace edition (deletion and insertion). Another way of formulating the error class is all 1-Levenshtein distance, plus 2-Levenshtein distance restricted to consecutive character edition.

*1) Character segmentation and recognition errors:* We built a noisy channel model to learn how OCR corrupts single characters or character segments using a prior knowledge set.

In order to create such a model, both OCR tokens and ground truth tokens are aligned at a word level with their calculated primitive Levenshtein distance. Out of this set we keep the alignments from the error class described above, and issue for each a segment correction instance. For example, given the aligned pairs (tlne, the), (amual, annual), their issued correction instances are: ln ⟶ h, m ⟶ nn. Subsequently, these instances are aggregated to a weighted confusion matrix.

[2]Based on modified Levenshtein distance where further primitive edit operations (character merge and split) are used, also known as 2:1 and 1:2 alignments.

TABLE I
EXAMPLE OF THE CORRECTION CANDIDATES GENERATION

| Correction candidates | tlna | grael | wollof | Chima |
|---|---|---|---|---|
| | the | great | walof | Cbina |
| | tlme | greet | wall of | Chna |
| OCR text | tlne | graat | wallof | China |

The limitation to primitive 1-Levenshtein distance plus 2:2 alignments corrections was based upon their relative ease of generalization, implied by the error class high proportion and recurrence in the erroneous word set, as well as its ease of implementation.

*2) Word segmentation errors:* Segmentation errors, also known as spacing errors, occur when a whitespace character is omitted between two words or erroneously inserted between two characters in a single word. This error class could not be generalized to the test data with the noisy channel approach, as segmentation errors are much more affected by text alignment and fonts than by preceding and following characters. This result was inferred by the relatively frequent single occurrences of such segment corruption and correction pairs. Therefore, we generate space omission correction candidates by joining two consecutive words into a single one and validating it on a unigram dictionary, and space insertion correction candidates by combinatorially splitting long words (more than 5 characters on our Arabic dataset) into two different words and validating them on a bigram dictionary.

This method ignores some types of errors, for example multi-error misspellings or m:n alignment error but avoids correction overfit, namely, these cases cannot be generalized to test data and are typically less frequent on our dataset.

We use the confusion matrix and word split/joins to expand a tokenized word into its possible corrections, forming all together the correction-candidates vector. The candidate generation is rule-base, where every character segment in a word is looked up in the confusion matrix and replaced by a possible segment correction, and every word is split or joined with its following word and looked up in the dictionary.

An example of the generation process can be seen in Table I.

### D. Candidate Ranker

The ranker's role is to produce an ordered word vector of correction candidates, calculating a score for each correction candidate, which correlates with how probable a correction is at a specific position. Every candidate is scored independently from all others in the word vector; then this candidate is compared to all the other correction-candidates. This stage does not take into account the original OCR output, as it has different features and will be considered in a secondary stage.

As a preliminary stage, the input vector was cleaned from all its non-dictionary words. As the dictionary is based on a large corpus, this procedure has only a negligible deleterious effect, while throwing away a considerable amount of irrelevant candidates and facilitating the scoring task.

In a secondary stage the word score is calculated by a trained regression model using the word's features as input.

*1) Feature Extraction:* The features were extracted at the word level:

- Confusion weight – The weight attribute of the corruption-correction pair in the confusion matrix, which is the number of occurrences of this pair calculated by the noisy channel on the training set. This feature reflects the OCR error model in accordance with the OCR engine performances over the document images training set, generally affected by font characteristics and scan quality.
- Unigram frequency – The unigram document frequency, providing a thematic domain and language feature independent of adjacent words or document context.
- Backward/Forward bigram frequency – The maximal document frequency of the bigram formed by a correction candidate and any candidate at the preceding/following position. This feature is valuable as it contains an intersection between language model and domain context, but is non-existent for many of the bigrams and is redundant if one of the unigrams does not exist. Although the bigrams should have been calculated in comparison to all the correction candidates, it was taken only on the OCR output due to calculation complexity and the relative rarity of sequential word errors. Furthermore, we set a cutoff frequency to overcome performances issues in the extraction stage.

No subsequent normalization procedure had to be made in order to linearize the feature effect for later linear regression modeling. In other words, the confusion weight behaves linearly, as well as the term frequency features that proportionally promote frequent corrections relative to their appearance in a similar corpus. Table II demonstrates the candidates feature extraction result.

*2) Ranker:* The ranker was trained from the OCR erroneous word set. Note that this set comprises solely words with extended single-error misspellings, words that the candidate generator supposedly generates. We used the training words to generate their correction-candidates vector and with their extracted features, with the single correct candidate marked with a positive output, as can be seen in Table II.

The appending of these vectors created a large training set used to create a regression model that attributes a continuous score to every correction candidate. This model was used to rank the correction-candidate vector and to sort it in descending order.

The choice of a ranker over a classifier was made to permit further applications of the ranked vector, such as outputting several words for information retrieval purposes or using them in a secondary correction process as we did. The scores could also be used to evaluate the process itself or to expose the correction confidence to the user.

*E. Correction Decision Maker*

The correction decision maker is a classifier that decides whether a replacement should be made of the OCR word with its highest ranked correction-candidate. Such a replacement is made in case the candidate is more likely to be the correct word at this position, as represented in Table III. We will refer to the OCR word and its highest-ranked correction-candidate as an "correction pair".

*1) Feature Extractor:* The decision is calculated by a trained regression model using the correction pair features as input:

- Confidence – An OCR output metric at a character level, which is generalized to a word level by taking the minimal confidence of the characters forming the word.
- Term frequency – The term frequency in the document, calculated by its frequency in the OCR text. This gives document level contextual information, as words forming a document tend to repeat themselves. A common problem of this feature is its bias to consistent OCR mistakes, thus it must be dealt with precaution.
- Proportional dictionary features – The same feature as used above. The proportion metric was included in order to adapt the features to comparative features that have a linear sense. A simple smoothing method was used to handle null-occurrences.

*2) Decision Maker:* The correction decision is made by a model trained on the total transcribed corpus of correction pairs. Pairs with erroneous OCR word and correct candidate were marked with a positive output, as shown in Table III, indicating that these cases are suitable for replacement.

## IV. Testing the Model

The model was tested on 50 articles containing a total of 15,000 words. The evaluation of the method was done by a ceiling analysis to understand the performance of every phase independently, as well as a conclusive evaluation for the entire process.

*A. Correction-Candidate Generation*

*1) Error Distribution:* On the test set, 80% of the erroneous words belong to the 1 primitive Levenshtein (including whitespace) misspellings. This observation gives an upper limit on the improvement ability of this method. About half of these misspellings were character substitution, 30% character deletion or insertion, 10% 2:1 and 1:2 alignments and 10% spacing errors. Pushing this upper limit upwards would require better image processing, for example binarization, layout analysis and word detection, or improving the character-level recognition.

*2) Retrieving the Correct Word:* Out of the misspelled words, 74% had been retrieved in their correct spelling in the correction-candidate generation process, suggesting that the OCR errors belong to a wider error set that the one trained on. This fair result can be attributed to the text variability of the archive we experiment on. Optimizing this result could be acquired by enlarging the training set or by generation candidates by an additional fuzzy logic against a dictionary or a trigram dictionary to retrieve possible word completion. A typical 4–5 character long word had up to 110 different correction-candidates, complicating the task of selecting the correct one.

TABLE II
EXAMPLE OF A TRAINING VECTOR FOR THE OCR WORD "GRAAT"

| Candidates | Confusion weight | Unigram frequency | Backward bigram frequency | Forward bigram frequency | Output |
|---|---|---|---|---|---|
| graat -> great | 41 (a->e) | 17,222 (great) | 1,238 (the great) | 73 (great wall) | 1 |
| graat -> greet | 5 (aa->ee) | 3,124 (greet) | 27 (the greet) | 0 (greet wall) | 0 |

TABLE III
A SCHEMATIC EXAMPLE OF A CORRECTION DECISION TRAINING OBSERVATION

| Correction pair (OCR word, top candidate) | Inverse proportions | | | | OCR confidence | Confusion weight | Decision |
|---|---|---|---|---|---|---|---|
| | unigram | backward bigram | forward bigram | term frequency | | | |
| (graat,great) | 100 | 10,000 | 20,000 | 500 | 0.4 | 15 | 1 |

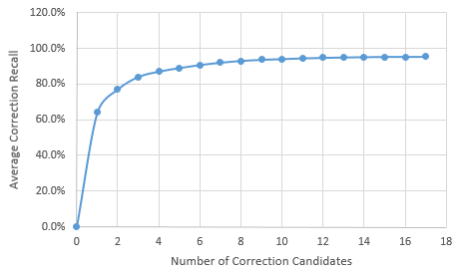

Fig. 1. Average recall on erroneous words as function of correction-candidates

### B. Candidate Ranker

The preliminary dictionary lookup qualification stage leaves on average 30 candidates to rank for a typical word as above.

We tried various ranking techniques on our training set and validated the results using $k$-fold training. A logistic regression model outperformed other models, yielding the results shown in Figure 1.

Calculated for words that have a valid candidate, the best model is able to find the proper correction within the top 5 proposed candidates for 90% of the words, and within the highest ranked candidate for 64% of the words. By the sparsity of the bigram feature we can attribute the mediocre result to the out-of-domain language model. Improving this result demands better features, for example, specifying that the corpus-based dictionary belong to the same thematic domain, or expanding the training set in order to enhance the confusion feature.

### C. Correction Decision Maker

Table IV reports the decision model performance, ignoring the decision on erroneous words that did not have their correct candidate highest ranked. The critical factor in this stage is the false positive rate, namely rejecting a correct OCR word in favor of its correction-candidate, as most of OCR words are correct and such rejections would significantly harm the reliability of the method. Therefore, the trained model gives preference to false positive rate diminution over false negative diminution. The main reason for these significant results is the bigram proportion feature, which in case the OCR word was invalid resulted in a very high number, as coincidental words rarely follow in text.

TABLE IV
PERFORMANCE OF THE DECISION MODEL FOR WORD CORRECTION

| | OCR word is actually | |
|---|---|---|
| | correct | incorrect |
| Reject OCR word | 2% | 94% |
| Accept OCR word | 98% | 6% |

### D. Overall Results

The baseline OCR text WER on the test collection is 30% at document-level, implying a relatively hard OCR task. Applying the presented method on the data improved the measure by relative 30% resulting in a 21% WER, which is a considerable improvement on difficult data. Minor changes, such as data cleaning and model optimization, should significantly improve the WER measure on this collection, and should compete with state-of-the-art OCR error spelling methods presented in the background.

The ceiling analysis clearly designates the ranker as a weak link, being apparently based on inaccurate dictionary features. The current ranking model gives a sufficient result for the top 5 ranking, but the method design requires the correct candidate to be the highest ranked, a task for which the current features quality is insufficient. The correction-candidate generation, trained by a noisy channel, misses a quarter of the corrections on the test because of the error types variability, suggesting this should be improved by additional correction candidate generation methods that do not rely solely on training errors. The correction decision maker is effective; with its large training set and indicative features one can expect similar results for different datasets.

## V. CONCLUSIONS

This paper has examined the use of machine-learning techniques for improving OCR accuracy by using the combination of a number of features to correct misspelled OCR words. The relative independence of the features, issuing from language model, OCR model and document context, enable a reliable spelling model that can be trained for most languages and domains. Although the results of the experiment on Arabic OCR text are only preliminary, they already impose a lower-limit on final improvements, showing an improvement in accuracy for every additional feature, implying the superiority of our multi-feature approach over traditional single-feature approaches. We can infer from the bigram feature significance that the contextual word completion is a reliable method for a machine as well for the human eye. Lastly, given a large out-of-domain corpus to extract a correction dictionary and to train a language model is a fine strategy for correcting a morphologically rich language such as Arabic with a 35% reduction in word error rate. An adaptation of this model to one that permits multi-word choices at a single textual position can be applied to information retrieval tasks to improve their recall even further.

For future work, correction-candidate generation and ranking improvements are to be considered as implied by the ceiling analysis. The rule base correction-candidate generation could be replaced by an unsupervised process, for example a dictionary-based expansion using a fuzzy unigram logic or a "gap 3- gram" to give generate correction candidates based on conditional left and right neighbors [8]. An improvement in ranking could be achieved by building a domain-specific language model, giving a considerable gain in strength for the language dependent features. Alternatively, one could determine the size of a sufficiently large corpus to generate a desirable confusion matrix.

A complete evaluation of this method on the Arabic Press Archive of the Moshe Dayan Center at Tel Aviv University is in preparation. This dataset has a simpler page layout and a more consistent writing format than the current set, suggesting its OCR errors will turn out to be more consistent. Moreover, large in-domain corpora are available and can be used to produce more accurate features hence may significantly improve ranking and classification performance.